# Requirements Issues in SoC and SoS

## Dagstuhl Seminar 12442

### Proposition Paper

Andrea Zisman

Department of Computer Science
City University London
Northampton Square, EC1V 0HB, UK
a.zisman@soi.city.ac.uk

---

The complexity of current computer-based applications, the rapid changes in market conditions and regulations, the dynamic creation of business alliances and partnerships, and the need to support the changing demands of users, require software systems to be more flexible, adaptable, and versatile. In order to support the above requirements, we have been facing the proliferation of software systems that are developed based on the composition of other already existing systems or services. Examples of such systems are found in energy efficiency and management systems, transport and traffic management systems, control systems for food and water supply chains, and Internet of Things applications.

Service-oriented computing (SOC), a new paradigm that envisages software as a temporary service rather than permanent property, aims to provide a more flexible approach to software development. In this context, services are loosely-coupled autonomous computer-based entities owned by third parties and representing different functionality, which can be combined to realise applications and create dynamic business processes. An essential characteristic of the SOC paradigm is the rapid use of such services to develop low-cost and easily composed distributed applications. Similarly, systems-of-systems (SoS) formed as a composition of software systems that are discovered, selected, and composed during runtime to create a more complex system also provides a more flexible and adaptable approach to software system development. The software systems in SoS applications need to fulfil not only the purpose of the (global) SoS application, but they also need to fulfil their own (isolated) purposes and to continue to do so even when they are not associated with a global application.

In both service-oriented computing and systems-of-systems applications their constituents services and systems (components) are developed by different entities, independently, and without considering the fact that they may be integrated with other components to support different scenarios, missions, and capabilities. Moreover, the environments in which these applications are executed, and their requirements, are partially known at design-time. In such settings, new unpredicted situations will emerge at run-time. One of the main challenges is how to develop these applications in a way that they can support new requirements while trying to fulfill their goals and objectives.

More specifically, the engineering processes and activities for developing software systems based on the composition of services, systems, or both is complex and require the need to support (i) modeling activities among various complex systems, (ii) integration of components that may be in different stages of their development and evolution, (iii) coordination of missions, capabilities, and requirements across the various services and systems, (iv) emergent unpredicted behavior during run-time, and (v) the use of the same components (services and systems) by different applications with distinct goals and objectives. Moreover, it is necessary to provide ways to support autonomous and automatic adaptation of these applications so that they can continue their operation despite changes in or emergence of new requirements, changes in the context and use of the applications and their components, or even failures or changes in their services and systems. Furthermore, it is also important to support adaptation of the services and systems themselves due to changes and new demands of the applications, users, and environment.

The characteristics and challenges of the above applications call for changes in the requirements engineering process. More specifically, it is necessary to deal with the fact that in these applications requirements are partially known at design-time. Furthermore, requirements engineering should be seen as a bi-lateral activity in which it is necessary to consider what the applications to be developed need to provide, as well as what the existing candidate services and systems that will form the applications can (or are prepared to) provide. It is important to have ways of dealing, during both design and run-time of the applications, with (a) the limitations that the available services and systems may cause with respect to the overall goals and objectives of the applications, (b) the potential that the existing available services and systems may be able to provide that may have not been initially thought by developers, and (c) the inconsistencies that may exist between the goals and objectives of the applications and their existing candidate constituent services and systems. Moreover, requirements engineering needs to deal with uncertainty and incompleteness of the available services and systems.

The elicitation of the requirements needs to be executed in an iterative way in which requirements will change and evolve depending on what the existing services and systems can provide. Furthermore, existing services and systems need to be described and represented in a clear and unambiguous way to facilitate identification and use of these services and systems by the applications, in an automatic way during run-time of the applications, when they can fulfill the requirements of the applications. It is also essential to provide support for negotiation of requirements between applications and their constituent services and systems.

Another issue is concerned with the situation in which there are no available candidate services or systems to fulfill the requirements of the applications during run-time. In such scenario, it is possible that the lack of candidate services or systems to fulfill the requirements of the application may demand changes in the requirements themselves (i.e., requirements adaptation). For example, some requirements may need to be relaxed. It is, therefore, necessary to develop techniques and approaches to manage adaptation and negotiation of the application requirements during run-time. Techniques are also necessary to verify that changes in the requirements will not cause inconsistencies or discrepancies in other parts of the applications and their components.

One other challenge is concerned with the need to improve the quality of the services and systems in order to allow differentiate themselves from other similar services and systems, and to react dynamically to market changes and user demands. This improvement in quality generates new requirements for the services and systems that may conflict with the requirements of the applications already using them. It is essential to create new techniques and approaches to manage requirements dependencies and conflicts that may exist between the services and systems, and between the applications and their participating services and systems.

Service-oriented computing and systems-of-systems applications involve cross-cutting domains. It is possible to have systems created for different domains (e.g., media, banking, transport) that will be used by an application incorporating these systems. Therefore, new ways of dealing with requirements that are not anymore domain specific, but that involve several domains at the same time are needed.

Some approaches have been proposed to support requirements engineering in adaptive software systems [1][2][4]. These approaches range from new languages to address uncertainty when specifying the behaviour of adaptive systems [6][10], to requirements monitoring frameworks for verification of violations of system's properties [3][7][9], and changes in requirements due to other requirements and certain conditions [8]. Other approaches were proposed to support multi-layered monitoring and adaptation of service compositions [9][11] and non-intrusive monitoring [5]. However, despite some advances made in the area, it is still necessary to develop new processes, techniques, methodologies, and tools to support requirements engineering in the scope of adaptable service-oriented computing and systems-of-systems applications.

In summary, the characteristics of service-oriented computing and systems-of-systems applications generate several research challenges with respect to design and run-time elicitation, management, representation, and fulfillment of requirements that need to be addressed. Existing requirements formalisms, techniques, approaches, processes, and tools are not adequate to support the issues that are present in these new classes of software systems. Together, the academic and industrial communities need to provide ways of dealing with these new challenges.